\documentstyle[11pt]{elsart}    
\input psfig
\def\ltsim{\raise 2pt \hbox {$<$} \kern-1.1em \lower 4pt \hbox {$\sim$}}
\def\ltapprox{\raise 2pt \hbox {$<$} \kern-1.1em \lower 5pt \hbox {$\approx$}}
\def\gtsim{\raise 2pt \hbox {$>$} \kern-1.1em \lower 4pt \hbox {$\sim$}}
\def\gtapprox{\raise 2pt \hbox {$>$} \kern-1.1em \lower 5pt \hbox {$\approx$}}
\def\arcsec{$^{\prime\prime}$}
\def\arcmin{$^{\prime}$}
\def\degrees{$^{\circ}$}
\def\etal{{\it et al.~}}

\def\skuno{\vskip 20pt}

\def\p0{\phantom{0}}

\def\ph1{\phantom{1}}
\begin{document}
\begin{frontmatter}

\title
{Halo and Relic Sources in Clusters of Galaxies}
\skuno
\skuno
\author[fis]{G. Giovannini\thanksref{g}},
\author[ira]{L. Feretti\thanksref{l}}

\address[fis]{
Istituto di Radioastronomia -- CNR, via Gobetti 101, I--40129
Bologna, Italy.
Dip. Fisica, Univ. Bologna, 
Via Berti-Pichat 6/2, I--40127 Bologna, Italy.}
\address[ira]{
Istituto di Radioastronomia -- CNR, via Gobetti 101, I--40129
Bologna, Italy.}

\thanks[g]{ggiovann@ira.bo.cnr.it}
\thanks[l]{lferetti@ira.bo.cnr.it}

\begin{abstract}

New images of 7 radio halos and relics, obtained with the Very Large Array
at 20 or 90 cm, are presented here. The existence of a cluster-wide
radio halo in the clusters A~665 and CL~0016+16 is confirmed. Both these
clusters share  the properties of the other clusters with radio halos, i.e.
are luminous in X-rays,  have high temperature, and show recent
merger processes. 

No diffuse sources are detected in a sample of clusters showing 
at least a tailed radio galaxy within 300 kpc from the cluster center, 
indicating that the connection between tailed radio galaxies and halos
 is not relevant. For these clusters we give limits to the surface brightness
and to the angular size of possible undetected diffuse sources.


\end{abstract}


\end{frontmatter}

\vfill\eject
\section{Introduction}

Diffuse radio sources in clusters of galaxies are the probe of the
existence of non-thermal emission from the  cluster intergalactic 
medium (IGM). 
These sources have no optical identification, steep radio spectrum and
size of $\sim$ 1 Mpc. Diffuse sources with regular shape, 
permeating the cluster center are called
{\it radio halos}, the prototype being the radio halo Coma~C in the
Coma cluster (Giovannini \etal 1993). 
Other  similar sources, with irregular morphologies,
located in cluster peripheral regions are named {\it relics}.
 In a few clusters,  both a central halo and a peripheral 
relic are present.

The difficulty in explaining these diffuse sources
arises from the combination of 
their large size and the short lifetimes of radiating electrons.
Several theoretical models have been proposed: (i) in-situ
acceleration by turbulent gas motion or 
 shocks produced in the intergalactic medium  during  
the merger of a subgroup into the main cluster,
(ii) diffusion of relativistic
electrons out of the present head-tail radio galaxies, (iii) secondary
particle production by hadronic interaction of relativistic
protons with the background gas of the IGM.  

The knowledge of the physical properties of these sources, and
of their origin and
evolution is limited by the low number of halos and relics well studied
up to now.
To  search for new sources of this class
and to test the possible correlation between the
presence of tailed radio galaxies and diffuse sources suggested by Giovannini
\etal (1993), we have obtained observations
with the Very Large Array (VLA
\footnote {The National Radio Astronomy
Observatory is operated by Associated Universities, Inc., under contract
with the National Science Foundation.}) of 21 clusters with the
following selection criteria:
a)  10 clusters where the possible presence of a diffuse 
source was suggested in the literature,
but the angular resolution and sensitivity of the available images was
not good enough to confirm its existence and derive the source parameters;
b) 11 clusters showing at least a tailed radio 
galaxy within 300 kpc (0.1 Abell radius) from the cluster center
in  the sample of O'Dea and Owen (1985). 


To properly map these extended sources one needs
large  sensitivity to the
extended features, but also with the high resolution 
necessary to distinguish a real halo from the blend of unrelated 
sources.
Observations were carried out using different VLA configurations  at 20 cm
for more distant clusters whereas the nearest clusters
were observed at 90 cm.
We present here sensitive images of the detected diffuse sources.
For the computation of intrinsic parameters, 
a Hubble constant H$_0$=50 km s$^{-1}$ Mpc$^{-1}$ and a deceleration 
parameter q$_0$=0.5 are assumed. 

\section {Observations and data reduction}

The list of observed cluster of galaxies is given in Table 1 where the 
observing frequency and the VLA configurations are also given. 

The bandwidths in the 1.4 GHz observations are of 25 MHz for the B array, 
and of 50 MHz for the C and D arrays. At 0.3 GHz, the bandwith is
of 3.125 MHz.

The observing frequency and array were chosen in order to have a good 
sampling of  short spacings, ensuring that a halo-type source similar to 
Coma~C could be easily detected and imaged.
We note that Coma~C  is completely resolved out from the VLA at 20 cm also
in its more compact array (D), because of missing short spacings. 
For the most distant clusters in the present sample, the VLA in
D configuration at 20 cm was found to be suitable, whereas for the 
nearest sources  observations at 90 cm were  necessary.  The
signal to noise ratio achieved by our observations should be enough to map
halo sources with a surface brightness 3-5 times lower than Coma~C.
Moreover, each cluster was observed with higher angular resolution 
to allow the separation of the  discrete sources from 
the diffuse cluster emission.

\begin{table}
\caption{Observed clusters of galaxies}
\begin{flushleft}
\begin{tabular}{llllll}
\hline 
\noalign{\smallskip}
Name  & z  & $\nu$ & Array & Int. time & Selection  \\
      &    & GHz  &       & min       &       \\
\noalign{\smallskip}
\hline
\noalign{\smallskip}
A~84  & 0.103  & 1.4   & B,C,D  & 60  & 1   \\
A~85  & 0.0555 & 0.3   & B,C    & 40   & 2   \\
A~119  & 0.0441 & 0.3   & B,C    & 40   & 1   \\
A~401  & 0.0739 & 0.3   & C      & 20   &  3  \\
A~610  & 0.0956 & 1.4   & C,D    & 170   & 4  \\ 
A~629  & 0.138 & 1.4   & C,D    & 40   & 1   \\
A~665  & 0.1818 & 1.4   & C,D    & 240   & 5 \\ 
A~754  & 0.0542 & 0.3   & B,C    & 40   & 3  \\
A~1132 & 0.1363 & 1.4   & C,D    & 40   & 1   \\
A~1190 & 0.0794 & 0.3   & B,C    & 40   & 1   \\
A~1314 & 0.0341 & 0.3   & B,C,D  & 60   & 1   \\
A~1609 & 0.113 & 1.4   & C,D    & 40   & 4   \\
A~1775 & 0.0696 & 0.3   & B,C    & 40   & 1   \\
A~2142 & 0.0894 & 1.4   & C,D    & 40   & 6   \\ 
A~2218 & 0.1710 & 1.4   & C,D    & 40   & 5   \\
A~2220 & 0.1106 & 1.4   & C,D    & 40   & 1   \\
A~2250 & 0.0654 & 0.3   & C      & 20   & 1   \\
A~2289 & 0.2276 & 1.4   & C,D    & 40   & 1   \\
A~2572 & 0.0384 & 0.3   & C,D    & 40   & 1   \\
CL~0016+16 & 0.5545  & 1.4  & B,C,D  & 200   & 5  \\
0917+75 & 0.125  & 1.4  & C,D    & 40    & 7 \\ 
\noalign{\smallskip}
\hline
\label{olog}
\end{tabular}
\end{flushleft}
\par\noindent
{\em Selection criterion.} 1: cluster containing at least one tailed
radio galaxy at the cluster center; 2: Joshi \etal 1986 (relic); 
3: Harris \etal 1980 (halo); 4: Valentijn 1979 (halo); 5:
Moffet \& Birkinshaw 1989 (halo);  6: Harris \etal 1977 (halo); 
7:  Harris \etal 1993 (halo).
\end{table}

The data were calibrated and reduced with the 
Astronomical Image Processing System (AIPS), following the
standard procedure:   Fourier-Transform, Clean and Restore.
Self-calibration was applied to minimise the effects of
amplitude and phase fluctuations. 
At each frequency, the data of different  arrays were combined to 
obtain the final maps. Images were produced at different angular resolutions
to study both the extended structures  and the discrete sources.
 In the caption of each figure, we give the Half Power Beam 
Width (HPBW) and the noise level of the maps.
A comparison with available maps
from the NRAO VLA Sky Survey (NVSS, Condon \etal 1998) and the
Faint Images of the Radio Sky at Twenty-cm  (FIRST, Becker \etal 1995) 
Survey  was also made.

\section {Results}

Among the 10 clusters where the presence
of a diffuse sources was suggested by previous authors, 
we have detected diffuse sources in 7 clusters.
The parameters
of these sources are given in Table 2. We have not detected  halos or
relics in the sample of clusters selected because of the presence
of tailed radio galaxies. 
In Table 3 we give the surface brightness limit for the 
undetected extended  sources and the largest size detectable
in our observations. In fact, we cannot exclude the presence of a source
with size larger than the limit, because of the missing short spacings. 
Notes on the clusters hosting diffuse sources, and on some other interesting 
clusters are given in the following.
We present images of  all the diffuse sources,  as well as
a few maps of extended and/or peculiar radio galaxies
detected in this study. 

\begin{table}
\caption{Clusters with a diffuse source }
\begin{flushleft}
\begin{tabular}{ccccccccc}
\hline 
\noalign{\smallskip}
Name &  $\nu$ & S & Log P$_{1.4}$  & L.A.S. 
& L.L.S. & M.L.S.  & Type \\
     & GHz &   mJy  &   W/Hz  &  \arcsec   & kpc   & Mpc \\
\noalign{\smallskip}
\hline
\noalign{\smallskip}
A~85        &  0.3 & 2739   & 23.97$^1$ & 350 & 510   & 2.64 & Relic \\
A~610       &  1.4 & 18.6   & 23.88     & 185 & 438   & 2.13 & Relic \\
A~665       &  1.4 & 43.1   & 24.82     & 600 & 2360  & 3.55 & Halo  \\
A~2142      &  1.4 & 18.3   & 23.82     & 120 &  270  & 2.01 & Halo \\
A~2218      &  1.4 &  4.7   & 23.81     & 130 &  490  & 3.39 & Halo  \\
CL~0016+16  &  1.4 &  5.5   & 24.95     & 150 & 1100  & 6.66 & Halo \\
0917+75     &  1.4 & 100.6  & 24.86     & 520 & 1530  & 2.66 & Relic \\
\noalign{\smallskip}
\hline
\label{olog}
\end{tabular}
\end{flushleft}
\par\noindent
{\em Caption.} Col. 1: cluster name; Col. 2: observing
frequency; Col. 3: total flux density at the previous frequency, 
after subtraction of discrete sources; Col. 4: logarithm of radio power 
at 1.4 GHz; Col. 5: Largest Angular Size measured on the radio images
as the largest extension of the 2$\sigma$ level, deconvolved by the
HPBW; Col. 6: Largest Linear Size corresponding to the largest angular size;
Col. 7: Maximum Linear Size detectable with the available interferometric
observation; Col. 8: type of diffuse source.
\medskip
\par\noindent
{\em Note:} $^1$ Radio power at 1.4 GHz has been computed assuming a spectral 
index = 2.5 (flux density at 1.4 GHz = 68.4 mJy)
\end{table}

\begin{table}
\caption{Cluster without diffuse source }
\begin{flushleft}
\begin{tabular}{lccrcll}
\hline 
\noalign{\smallskip}
Name &  $\nu$ & HPBW              & $\sigma_{noise}$ & M.L.S. \\
      & GHz  &  \arcsec $\times$ \arcsec   & mJy/b &  Mpc    \\
\noalign{\smallskip}
\hline
\noalign{\smallskip}
A~84 &   1.4       & 40$\times$40      & 0.1   &   2.23       \\
A~119 &  0.3       & 60$\times$55      & 5.0   &   2.14       \\
A~401 &  0.3       & 65$\times$62      & 2.5   &   3.41       \\
A~629 &  1.4       & 49$\times$40      & 0.2   &   2.88        \\
A~754 &  0.3       & 109$\times$88     & 6.5   &   2.59       \\
A~1132&  1.4       & 54$\times$38      & 0.14  &   2.85        \\
A~1190&  0.3       & 77$\times$71      & 6.8   &   3.63        \\
A~1314&  0.3       & 188$\times$153    & 2.7   &   3.93        \\
A~1609&  1.4       & 43$\times$36      & 0.18  &   2.45        \\
A~1775&  0.3       & 72$\times$63      & 5.1   &   3.24        \\
A~2220&  1.4       & 48$\times$35      & 0.16  &   2.41        \\
A~2250&  0.3       & 74$\times$69      & 5.2   &   3.06        \\
A~2289&  1.4       & 52$\times$35      & 0.20  &   4.15        \\
A~2572&  0.3       & 190$\times$182    & 2.4   &   4.39       \\
\noalign{\smallskip}
\hline
\label{olog}
\end{tabular}
\end{flushleft}
\par\noindent
{\em Caption.}  Col. 1: cluster name; Col. 2: observing
frequency; Col. 3: Half Power Beam
Width of the observation;  Col. 4: noise level; 
Col. 5: Maximum Linear Size detectable with the available interferometric
observation. 
\end{table}

\subsection {Notes on individual clusters of galaxies with diffuse sources}

\noindent
{\bf A~85}. This cluster was studied in detail by Bagchi \etal (1998). 
In Fig. 1
we present the image of the central cluster region obtained at 90 cm,
with angular resolution of 25\arcsec.
A small size radio emission is detected from the central cD galaxy (A),
and two head-tail radio galaxies are visible (B,C).
 
Moreover, two extended regions of radio emission with no obvious
optical identification are present: they are labelled 'R' and 'D'
in Fig. 1. Source 'R' is the well known relic source of steep spectrum 
studied by Joshi \etal  (1986) and Bagchi \etal (1998).
In this image, the relic has an irregular  structure.
An image with slightly better angular resolution (see Fig. 2) shows 
that the relic  has a complex structure with evidence of 
internal substructures and filaments. This is confirmed by the high 
resolution image  presented by Andernach \etal (2000).
The total flux density (2.74 Jy) is lower than the total flux density measured
with the Ooty Synthesis Radio Telescope
 at the same frequency (3.15 Jy, Joshi \etal 1986). 
We refer to Bagchi \etal (1998)
for the total radio spectrum of this source.
We obtained a spectral index map between 90 and 20 cm,
combining our image with the 20 cm  image retrieved from the NVSS, 
presented in Giovannini \etal 1999.)
Because of the low resolution of the NVSS image (45$^{\prime\prime}$),
the substructures are smoothed in the maps used for the
spectral comparison.  The spectrum  in
very steep ($\alpha \sim$2.5 - 3.0)  with no evidence of substructure.

The sources `B' and 'D' are blended together in the 90
cm image by Bagchi \etal (1998), where the extended feature
formed by these two sources is suggested to be another possible
diffuse source. From the present high resolution image, we distinguish the
tailed radio galaxy 'B' and remark
the existence of the extended feature labelled 'D', which
has a steep spectrum  
($\alpha^{1.4}_{0.3} \sim 2-2.5$, obtained by comparison with the NVSS) 
and a total 
extent of about 2\arcmin, corresponding to $\sim$175 kpc. 
Its shape and the presence of a nearby galaxy with a possible faint
radio emission  at about the 3$\sigma$ level
suggests that this structure could be a dying
tailed radio galaxy, where the nuclear emission has almost completely ceased.

\noindent
{\bf A~610}. We detect a faint diffuse emission (Fig. 3), coincident with the
halo source detected by  Valentijn \etal (1979).
It is elongated in shape and displaced from the  
cluster center, thus we classify it as a  relic source.  
The high resolution image given in Fig. 4 confirms
that the source is diffuse.
Using the fux at 0.6 GHz from Valentijn \etal (1979), 
S$_{0.6~GHz}$ = 59 mJy, 
a spectral index  $\alpha^{1.4}_{0.6}$ = 1.4 is derived.

\noindent
{\bf A~665}. Moffet \& Birkinshaw (1989) 
detected a diffuse emission coincident with
the cluster center. Our image (Fig. 5) shows that this halo source 
is very extended and  elongated
in the SE-NW direction. 
The highest resolution map 
(not shown here) indicates that  
 2 faint discrete sources are present within the diffuse
radio  emission (crosses in Fig. 5)  and that the radio emission 
is asymmetric with respect to the cluster center,
being brighter and more extended toward NW.

\noindent
{\bf A~2142}. The presence of a halo in this cluster was
suggested by Harris \etal (1977). 
We confirm the presence of diffuse emission, located around the brightest
cluster galaxy  (Fig. 6), as also detected
in the NVSS (Giovannini \etal 1999).
The radio emission could be 
more extended in S - SW direction up to 27$^\circ$ 10\arcmin~ in declination
(2 sigma level), but 
more sensitive observations are necessary to confirm the reality of this 
feature. 
In the FIRST survey, the
diffuse emission is completely resolved out, and no
emission is detected at a flux  level of 0.4 mJy/beam
(3$\sigma$) from any of the the bright galaxies
present in the region of the extended radio emission.
The radio source could therefore be classified as a cluster halo.
We note, however, that this source is much smaller than radio halos commonly 
found in clusters. 
We cannot rule out the possibility that it is the remnant of
a single galaxy which was active in the past.

\noindent
{\bf A~2218}. We confirm the existence of the diffuse source found by Moffet \&
Birkinshaw (1989). It is slightly displaced from the cluster
center, toward East (Fig. 7). In a deep image obtained by 
Zwaan \etal (2000) the diffuse source is more
extended, with radio emission of very
low surface brightness  also to the West of the
cluster center.
This halo source is smaller than 
typical cluster-wide radio halos. It is similar, 
both in structure and in size, to
the halo in A~1300 (Reid \etal 1999). As in the case of
A~1300, the diffuse radio source is located in the same
direction as the extension of the cluster X-ray emission
detected by the ROSAT HRI (Markevitch 1997).

At 5 GHz, the flux density of the halo is reported to be 0.6 mJy
(Partridge \etal 1987). Comparison of this flux with the
present measurement at 1.4 GHz provides a spectral index $\alpha^{1.4}_5$
$\sim$ 1.6, which should however be taken cautiously since the two maps
at the two frequencies do not have matched  Fourier coverage.

\noindent
{\bf CL~0016+16}.  We confirm the presence of a diffuse emission near the 
cluster center found by Moffet \& Birkinshaw (1989). Unfortunaly,
our high resolution
observations (VLA configurations B and C) are strongly 
affected by interferences therefore
we cannot produce a higher resolution map. 
The diffuse radio emission (Fig. 8) is located at the cluster center,
and is extended more than 1 Mpc, therefore it is a typical
radio halo. We note the presence of a fainter radio emission west of the 
halo, at around RA$_{2000}$ = 00$^h$ 18$^m$ 25$^s$, DEC$_{2000}$ =
16\degrees~ 26\arcmin~ 48\arcsec, coincident with a galaxy group visible in
the DPSS (see Fig. 8).

\noindent
{\bf 0917+75}. This source was studied in detail by Dewdney \etal (1991) and
Harris \etal (1993).
Our image (Fig. 9) shows that the diffuse source is elongated in 
 the E-W direction, with flux density and morphology
in very good agreement with their data.

As discussed in the above mentioned papers,  this source cannot be
obviously associated with a well defined cluster of galaxies. 
In the region, there are the 3 clusters A~786, A~787 and A~762, 
at redshift larger than  0.1, belonging to the Rood Group of clusters
of galaxies \#27, but
the source is located at very large distance from their centers
(respectively 30\arcmin, 45\arcmin, and  44\arcmin, corresponding
to $\sim$ 5 Mpc, $\sim$ 8 Mpc and $\sim$ 8 Mpc). 
Thus, although the radio emission is reminiscent of the
giant radio relic in A~3667 (R\"ottgering \etal 1997), its very large
distance from the cluster center would favour the
hypothesis that this diffuse source is related to a possible
group near the source center (see the optical DPSS image in Fig. 9) or to
the supercluster. This makes this source quite unique.

\subsection {Notes on individual clusters without diffuse sources}

{\bf A~84}. An extended head-tail radio galaxy
is present in the cluster center region, as reported by Rudnick \& Owen 
(1976) and O'Dea \& Owen (1985). In our map, 
the tailed radio galaxies is detected
to a much larger extent than previously believed, and for the first
time a prominent bend of the tail is visible (Fig. 10).

Moreover, a faint extended radio emission is detected at about 2\arcmin~ SE
of the tailed radio galaxy. This feature has no obvious optical
counterpart, and is completely resolved out in
a higher resolution map. 
Its structure and the presence of a galaxy at the boundary  of the
radio emission might support the hypothesis that it is 
the remnant of a tailed radio galaxy, where the nuclear emission
has ceased, but it could also be a distant  
unrelated source with no optical counterpart.

\noindent
{\bf A~119}. This cluster was studied in detail by Feretti et al. (1999).
The present data at 90 cm confirm the absence of  diffuse emission.

\noindent
{\bf A401} - Our image at 90 cm shows the same features detected by  Harris 
\etal (1980) and Roland \etal (1981), in particular the two components 
A and B (see Fig. 11). On the contrary, the 
 faint  extended emission  detected near the central cD galaxy in the
NVSS (see Giovannini \etal 1999) at 20 cm is not seen here.
With the present data, the existence of diffuse emission in this cluster
remains quite dubious, as well as the classification of source B as a 
cluster halo. 

\noindent
{\bf A~629}. No diffuse emission is detected in this cluster, which is
characterized by the presence of an extended wide-angle-tail  radio galaxy
and of a head-tail radio galaxy at the center (see Owen \& Ledlow 1997). 

\noindent
{\bf A~754}. The  extended emission at the cluster center, reported 
by Harris \etal (1980) and also visible in the  NVSS image
(Giovannini \etal 1999) is at the limit of significance in our map 
at 90 cm. This implies
that this feature cannot have a  steep spectral index between 90 and
 20 cm ($\alpha^{20}_{90}$ $<$ 0.7) and
confirms the suggestion of Giovannini \etal (1999)
that the radio emission  might be the blend
of many bright cluster galaxies in this region, rather than a diffuse halo.
The diffuse emission detected in the NVSS in the eastern peripheral 
region (Giovannini \etal 1999) cannot be seen in the 90 cm image because
of the too low angular resolution. 

\noindent
{\bf A~1609}. The radio halo suggested by Valentijn (1979)
has not been detected in our image, which shows several radio sources.
The extended  radio galaxy  with 
a tailed structure (Fig. 12) is actually the blend of two tailed
radio galaxies (Rudnick, private communication).

\noindent
{\bf A~1775}. 
The head-tail radio galaxy in this cluster (B1339+266B)  shows a
very narrow  tail (Fig. 13), 
extended $\sim$ 8\arcmin~ (850 kpc), i.e. twice more than in the 
image of Owen \& Ledlow (1997). A similar head-tail radio galaxy,  but with
a shorter extent ($\sim$550 kpc) is found in the cluster A 2256.

\noindent
{\bf A~2250}. The head-tail radio galaxy near the cluster center 
(B1709+397) is much more extended than in the previous map by
O'Dea and Owen (1985)
($\sim$ 6\arcmin, corresponding to $\sim$ 600 kpc), 
but no diffuse halo is visible (Fig. 14).

\section {Discussion and conclusions}

The present paper increases the number of halo and relic sources
with detailed radio information.
The new data on the two clusters A~665 and CL~0016+16 are of great 
importance in the study of cluster-wide radio halos, since they represent
two more examples of giant sources of this class.
Both the halos in A~665 and in CL0016+16 are more powerful than the Coma halo.

The cluster A~665 has a bolometric X-ray luminosity of 4.2 10$^{45}$
erg s$^{-1}$ (Wu \etal 1999), a temperature of 8.3 keV (Wu \etal 1999)
and no cooling flow (White \etal 1997). 
From X-ray data, it is suggested to be in 
a postmerger state, with an asymmetric X-ray brightness distribution,
extended in the NW direction  (Markevitch 1996, Gomez \etal 2000). 
This is also the orientation of the radio structure
(Fig. 15 and Sect. 3), indicating a spatial correlation between the radio and
X-ray structure, as found in other clusters (see e.g. Feretti 2000).

The cluster CL~0016+16 has a bolometric X-ray luminosity of  2.81 10$^{45}$
erg s$^{-1}$ (Wu \etal 1999). Its X-ray emission has been studied by Neumann
\& B\"ohringer (1997)
 who obtain a temperature of 8.2 keV, conclude that the cluster
has not yet formed a steady cooling flow, and report strong
evidence of substructure which is possibly due to merging of a galaxy
group with the main cluster. 
The radio-X-ray overlay, presented in Fig. 16, shows that  the radio
halo is  located  at the cluster center, at the 
position of the X-ray peak.

In the clusters A~2142 and A~2218, diffuse halos of about 500 kpc in size, 
or less, are detected. 
 Both these clusters are characterized by signatures of recent merger 
events (Markevitch 1997, Markevitch \etal 1998).  
No cooling flow is detected in A~2218, whereas
the cluster A~2142 has a massive cooling flow (Peres \etal 1998) centered
on the diffuse radio source. The presence of a radio halo in a cooling 
flow cluster is quite uncommon (Feretti 2000). The fact that the halo
in A~2142 has a small size could be related with the presence of the cooling
flow.
Future deep observations of more clusters
of galaxies are necessary to understand if  these sources have a different
nature with respect to  the  giant cluster-wide halos, or if they
are the low power-small size tail of the same class of radio sources.

We have detected no radio halos in the 11 clusters showing tailed
radio galaxies at the cluster center. 
In the Coma cluster, Giovannini \etal (1993) suggested that
the  tailed radio galaxy NGC4869,
which is orbiting around the cluster center, could
be  responsible for the supply of relativistic 
electron radiating within the radio halo Coma~C. 
Also in the clusters A~2255 (Feretti \etal 1997a) and A~2319 
(Feretti \etal 1997b), tailed radio galaxies
are found within
the diffuse radio emission, or close to it, supporting the hypothesis
that the tailed radio galaxies could provide relativistic electrons to the
radio halo. In the sample of tailed radio galaxies presented by 
O'Dea and Owen (1985), there are 15 clusters with at least a tailed
radio galaxy within 300 kpc from the cluster center (Giovannini \etal 1993). 
These are the 11 clusters studied here, plus A~85, A~1656 (Coma),
A~2255 and A~2256. The last 3 clusters show a halo source, while  A~85
hosts  a relic radio emission (Sect. 3). For the
11 clusters observed here,
we only give a limit in size and brightness for the presence
of a possible extended source. 
The lack of detection of bright halos in these clusters  would imply
that the connection between halos 
and tailed radio galaxies seems not relevant. 
On the other hand,  we cannot exclude 
that these tailed radio galaxies are not really at the cluster
center but simply projected onto it, or that they are  crossing the
cluster center and not
orbiting at the cluster center, as needed to fuel a diffuse source.
From rotation measure studies of the sources in 
 A~119 (Feretti \etal 1999), where no halo is detected,
it was derived that 
the two  powerful tailed radio galaxies are genuinely located 
at the cluster center. We note, in addition, that 
no tailed radio galaxy is found to be present in A~665, hosting a cluster-wide
radio halo.
We conclude that the tailed radio galaxies could
be a  source of relativistic electrons, but they are not
a necessary ingredient for the halo formation. These results are in
line with the model developed by Brunetti \etal (2000a), who suggest
that relativistic particles radiating within the halos were injected
in the cluster volume by past major merger events, starbust and AGN
activity. 
Following a recent suggestion  by Brunetti \etal (2000b),
the relativistic particles deposited by the tailed radio galaxies could
account for the emission of extreme ultraviolet radiation.

 According to  Bliton \etal (1998), the clusters containing narrow-angle 
tailed radio galaxies are in general dynamically complex systems udergoing
merger events. Actually, structure and other merger signatures have been
detected in all the clusters of Table 3 (e.g. A~754, Henry \& Briel 1995; 
A~119, Markevitch \etal 1998) which have been 
adequately observed in optical or X-ray, except
for A~401, which is a rather unusual cD cluster with 
neither a merger activity nor a cooling flow. Therefore, we confirm
that  the presence of a recent merger is not 
sufficient for the formation of a radio halo.

Recent studies have shown that radio halos are found in 
merging clusters which are also characterized by high X-ray luminosity,
i.e.  high temperature and large mass
(Giovannini \etal 1999, Feretti 2000).
The halos in A~665 and CL~0016+16  are fully confirming the above result.
The connection between halos and mergers, and the presence of halos
in hot, massive,  X-ray bright clusters can be interpreted in the framework
of the two-phase model suggested by Brunetti \etal (2000a) and
successfully applied to Coma C.  
According to this model, relativistic particles produced in the early
cluster history, are reaccelerated by the energy supplied  from shocks 
and turbulence in a recent merger event.  Alternative 
models and the theoretical implications of non-thermal emission from the
intracluster medium are reviewed by En{\ss}lin (2000).

\vfill\eject
\begin{ack}

 We thank the referee, Dr. T. En{\ss}lin, for helpful comments.

This research has made use of the NASA/IPAC Extragalactic Database (NED)
which is operated by the Jet Propulsion Laboratory, Caltech, under contract
with the National Aeronautics and Space Administration.

We acknowledge partial financial support from the Italian
Space Agency (ASI).

\end{ack}

\vskip 1 truecm

{\bf References}

\par\medskip\noindent
Abell G.O., Corwin H.G. Jr., Olowin R.P., 1989, ApJS 70, 1
\par\medskip\noindent
Andernach H., et al., 2000, {\it The Universe at low Radio Frequencies},
Symposium IAU 199, Pune (India), in press
\par\medskip\noindent
Bagchi J., Pislar V., Lima Neto G.B, 1998, MNRAS 296, L23
\par\medskip\noindent
Becker, R.H., White, R.L., Helfand, D.J.,  1995, ApJ 450, 559
\par\medskip\noindent
Bliton M., Rizza E., Burns J.O., Owen F.N., Ledlow M.J., 1998, MNRAS 301, 609
\par\medskip\noindent
Brunetti G., Setti G., Feretti L., Giovannini G., 2000a, MNRAS
Submitted
\par\medskip\noindent
Brunetti G., Setti G., Feretti L., Giovannini G., 2000b, in preparation
\par\medskip\noindent
Condon J.J., Cotton W.D., Greisen E.W. Yin Q.F., Perley R.A.,
Taylor G.B., Broderick J.J., 1998, AJ 115, 1693
\par\medskip\noindent
Dewdney P.E., Costain C.H., McHardy I., Willis A.G.,
Harris D.E., Stern C.P., 1991, ApJS 76, 1055 
\par\medskip\noindent
En{\ss}lin T.A. 2000, In: {\it The Universe at low Radio Frequencies},
Symposium IAU 199, Pune (India), in press, astro-ph/0001433
\par\medskip\noindent
Feretti  L.,  B\"ohringer H., Giovannini G., Neumann D.,
1997a, A\&A 317, 432
\par\medskip\noindent
Feretti L., Giovannini G. B\"ohringer H.,
1997b, New Astronomy 2, 501
\par\medskip\noindent
Feretti L., Dallacasa D., Govoni F. Giovannini G., Taylor G.B.,
Klein U., 1999, A\&A 344, 472
\par\medskip\noindent
Feretti L., 2000, In: {\it The Universe at low Radio Frequencies},
Symposium IAU 199, Pune (India), in press, astro-ph/0006379
\par\medskip\noindent
Giovannini  G., Feretti L., Venturi T., Kim K.-T., 
Kronberg P.P., 1993,  ApJ, 406, 399
\par\medskip\noindent
Giovannini G., Tordi M., Feretti L., 1999, New Astronomy 4, 141
\par\medskip\noindent
Gomez P.L., Hughes J.P., Birkinshaw M, 2000, ApJ, in press, astro-ph/0004263
\par\medskip\noindent
Harris D.E., Bahcall N.A., Strom R.G., 1977, A\&A 60, 27
\par\medskip\noindent
Harris D.E., Kapahi V.K., Ekers R.D., 1980, A\&AS 39, 215
\par\medskip\noindent
Harris D.E., Stern C.P., Willis A.G., Dewdney P.E., 1993,
AJ 105, 769
\par\medskip\noindent
Henry J.P., Briel U.G., 1995, ApJ 443, L9
\par\medskip\noindent
Joshi M.N., Bagchi J., Kapahi V.K., 1986, In: 
{\it Radio continuum processes in clusters of galaxies}, NRAO,
p. 73
\par\medskip\noindent
Markevitch M., 1996, ApJ 465, L1
\par\medskip\noindent
Markevitch M., 1997, ApJ 483, L17
\par\medskip\noindent
Markevitch M., Forman W.R., Sarazin C.L., Vikhlinin A., 1998, 
ApJ 503, 77
\par\medskip\noindent
Moffet A.T., Birkinshaw M., 1989, AJ 98, 1148 
\par\medskip\noindent
Neumann, D.M., B\"ohringer H., 1997, MNRAS 289, 123
\par\medskip\noindent
O'Dea C.P., Owen F.N., 1985, AJ 90, 927
\par\medskip\noindent
Owen F.N., Ledlow M.J., 1997, ApJS 108, 41
\par\medskip\noindent
Partridge R. B., Perley R. A., Mandolesi N., Delpino F.,
 1987, ApJ 317, 112
\par\medskip\noindent
Peres C.B., Fabian A.C., Edge A.C., Allen S.W., Johnstone R.M., White
D.A., 1998, MNRAS 298, 416
\par\medskip\noindent
Reid A.D., Hunstead R.W., L\'emonon L., Pierre M.M., 1999, MNRAS 302,
571
\par\medskip\noindent
Roland, J. Sol, H. Pauliny-Toth, I. Witzel, A. 1981, A\&A 100, 7
\par\medskip\noindent
R\"ottgering  H.J.A., Wieringa M.H., Hunstead R.W., Ekers R.D.,
1997, MNRAS 290, 577 
\par\medskip\noindent
Rudnick L., Owen F.N., 1976, ApJ 203, L107
\par\medskip\noindent
Valentijn E.A., 1979, A\&AS 38, 319
\par\medskip\noindent
White D.A., Jones C., Forman W., 1997, MNRAS 292, 419
\par\medskip\noindent
Wu Xiang-Ping, Xue Yan-Jie, Fang Li-Zhi, 1999, ApJ 524, 22
\par\medskip\noindent
Zwaan M.A., van Dokkum P.G., Verheijen M.A.W., Briggs F., 2000, poster
presented to the conference {\it Gas and Galaxy evolution}, Socorro, NRAO,
in press 

\vfill\eject

\begin{figure}
\caption{Isocontour map at 90 cm
of the central region of A85 overimposed on the optical 
image from the Digitized Palomar Sky Survey (DPSS). A is the central cD galaxy,
B and C are cluster radio galaxies with head-tail 
morphology. R is the steep spectrum relic source while D is an extended
steep spectrum source probably related to the near cluster galaxy
(see text).
The HPBW is 25\arcsec; the
noise level is 1.1 mJy/beam. Contour levels are: 3, 5, 10, 30, 50, 70, 120, 150
mJy/beam.}
\label{A 85}
\end{figure}

\begin{figure}
\caption{Radio image at 90 cm of the relic source in A85, with  
HPBW of 20\arcsec.  The
noise level is 1.5 mJy/beam. Contour levels are: 5, 10, 20, 30, 40, 50, 70, 100
mJy/beam }
\label{A 85.2}
\end{figure}

\begin{figure}
\caption{ Isocontour map at 20 cm
of the central region of A~610 overimposed the optical 
image from the DPSS. The position of the cluster center given by 
Abell \etal (1989) is RA$_{2000}$ = 07$^h$ 59$^m$ 15.6$^s$, 
DEC$_{2000}$ = 27$^{\circ}$ 06\arcmin~ 48\arcsec.
The HPBW is 50\arcsec $\times$ 45\arcsec (PA=0\degrees); the
noise level is 0.15 mJy/beam. Contour levels are: -0.4, 0.4, 0.8, 1.5, 3, 
6, 12, 15, 25, 50, 100 mJy/beam. }
\label{A 610,1}
\end{figure}

\begin{figure}
\caption{ Isocontour map at 20 cm of the central region of A610 overimposed the optical 
image from the DPSS. 
The HPBW is 18.0\arcsec $\times$ 16.7\arcsec (PA=13.5\degrees); the
noise level is 0.14 mJy/beam. Contour levels are: -0.4, 0.4, 0.8, 1.5, 3, 
6, 12, 15, 50 mJy/beam. }
\label{A 610,2}
\end{figure}

\begin{figure}
\caption{ Isocontour map at 20 cm of the central region of A~665 
superimposed onto the optical image from the DPSS.  Crosses
mark two unrelated sources, of flux $\sim$ 1.2 mJy each.
The HPBW is 52\arcsec $\times$ 42\arcsec (PA=0$^{\circ}$); the
noise level is 0.065 mJy/beam. Contour levels are: -0.2 0.2 0.4 0.8 1.5 3 6 12 
25 mJy/beam. }
\label{A 665,1}
\end{figure}

\begin{figure}
\caption{ Isocontour map at 20 cm of the central region of A2142. 
The HPBW is 40\arcsec~ and the
noise level  0.15 mJy/beam. Contour levels are: -0.3, 0.3, 0.5, 0.7, 1, 2, 
3, 5, 10, 20, 30, 40, 60 mJy/beam. }
\label{A 2141,1}
\end{figure}

\begin{figure}
\caption{ Isocontour map at 20 cm of the central region of A~2218.
The  cluster dominant galaxy is indicated by an arrow.
The source to the north-east is a foreground S galaxy.
The HPBW is 35\arcsec~ the
noise level is 0.09 mJy/beam. Contour levels are: -0.2, 0.2, 0.3, 0.5, 
0.7, 1, 3, 5, 10 mJy/beam. }
\label{A 2218,1}
\end{figure}

\begin{figure}
\caption{ Isocontour map at 20 cm of the central region of CL~0016+16. 
The HPBW is 60\arcsec $\times$ 45\arcsec (PA=0\degrees); the
noise level is 0.09 mJy/beam. Contour levels are: -0.2, 0.2, 0.4, 0.8, 
1.5, 3, 6, 12, 25, 50, 100 mJy/beam. }
\label{A 0016,1}
\end{figure}

\begin{figure}
\caption{ Isocontour map at 20 cm
of the extended source 0917+75. 
The HPBW is 30\arcsec; the
noise level is 0.08 mJy/beam. Contour levels are: -0.2, 0.2, 0.5, 0.7, 1, 
1.5, 2, 3, 5, 7, 10, 30, 50 mJy/beam. }
\label{A 0917,1}
\end{figure}

\begin{figure}
\caption{ Isocontour map at 20 cm of the head-tail radio galaxy in A 84 overimposed the
optical image from the DPSS. The HPBW is 11.5\arcsec; the
noise level 0.03 mJy/beam. Contours are: 0.1, 0.3, 0.5, 1, 5, 10, 30, 50,
 70, 90 mJy/beam. }
\label{A 84}
\end{figure}

\begin{figure}
\caption{ Isocontour map at 90 cm of A 401 overimposed onto the optical 
image from the DPSS. 
The HPBW is 65\arcsec$\times$62\arcsec(PA=64\degrees). Contour levels are: 
-5, 5, 10, 20, 40, 80, 120, 200 mJy/beam. 
}
\label{A 401}
\end{figure}

\begin{figure}
\caption{ Isocontour map of the central region of A 1609 overimposed the 
optical image from the DPSS. The arrows indicate the two radio emitting
galaxies, which give raise to the extended tailed structure.
The HPBW is 12.5\arcsec
$\times$ 11.4\arcsec (PA=-14\degrees); the
noise level is 0.18 mJy/beam. Contour levels are: -0.3, 0.2, 0.5, 1.5, 2,
3, 5, 10, 20 mJy/beam. }
\label{A 1609}
\end{figure}

\begin{figure}
\caption{ Isocontour map at 90 cm 
of the head tail radio galaxy B1339+266B
in A~1775 overimposed the 
optical image from the DPSS. 
The HPBW is 72\arcsec
$\times$ 63\arcsec (PA=83\degrees); the
noise level is 5.1 mJy/beam. Contour levels are: 10, 15, 20, 30, 50, 70, 
100, 200, 500 mJy/beam. }
\label{A 1609}
\end{figure}

\begin{figure}
\caption{ Isocontour map at 90 cm of the tailed radio galaxy B1709+397
in A~2250 overimposed the 
optical image from the DPSS. The HPBW is 74\arcsec
$\times$ 69\arcsec (PA=-53\degrees); the
noise level is 5.1 mJy/beam. Contour levels are: -15, 15, 30, 50, 100, 200, 300, 500,
700, 1000 mJy/beam. }
\label{A 1609}
\end{figure}

\begin{figure}
\caption{Overlay of the contour radio image of A~665 onto the grey-scale
X-ray image obtained with the ROSAT PSPC }
\label{boo}
\end{figure}

\begin{figure}
\caption{Overlay of the contour radio image of CL~0016+16
onto the grey-scale X-ray image obtained with the ROSAT HRI}
\label{baa}
\end{figure}

\end{document}